# Structural characteristics and physical properties of neutron stars: theoretical and observational research

Collazos J. Alfredo[1]

[1] *Peter the Great St. Petersburg Polytechnic University, Politechnicheskaya 29, 195251, St. Petersburg, Russia*
[1] E-mail: jacollazos@utp.edu.co

**ABSTRACT**

Neutron stars are one of the most extreme objects in the universe, with densities that can exceed those of atomic nuclei and gravitational fields that are among the strongest known. Theoretical and observational research on neutron stars has revealed a wealth of information about their structural characteristics and physical properties. The structural characteristics of neutron stars are determined by the equations of state that describe the relationship between their density, pressure, and energy. These equations of state are still not well understood, and ongoing theoretical research aims to refine our understanding of the behavior of matter under these extreme conditions. Observational research on neutron stars, such as measurements of their masses and radii, can provide valuable constraints on the properties of the equation of state.

The physical properties of neutron stars are also of great interest to researchers. Neutron stars have strong magnetic fields, which can produce observable effects such as pulsations and emission of X-rays and gamma rays. The surface temperature of neutron stars can also provide insight into their thermal properties, while observations of their gravitational fields can test predictions of Einstein's theory of general relativity. Observational research on neutron stars is carried out using a variety of techniques, including radio and X-ray telescopes, gravitational wave detectors, and optical telescopes. These observations are often combined with theoretical models to gain a more complete understanding of the properties of neutron stars.

## 1. INTRODUCTION

Neutron stars are one of the most extreme objects in the universe, with a mass greater than the Sun but a diameter only about 20 km. They are formed by the gravitational collapse of a massive star, leaving behind a highly dense and compact core made up mostly of neutrons (P.Haensel & A.Y. Potekhin 2007).

The structural characteristics and physical properties of neutron stars have been the subject of both theoretical and observational research. Theoretical models predict that the matter inside a neutron star is incredibly dense, with a density that can be as much as several times that of an atomic nucleus. The pressure inside the star is so high that it can sustain the gravitational force that would otherwise cause the star to collapse into a black hole.

Observationally, neutron stars have been detected through their emission of X-rays, gamma rays, and radio waves. The study of pulsars, a type of rapidly rotating neutron star that emits a beam of electromagnetic radiation, has provided important insights into the structure and properties of neutron stars. The precise timing of pulsar signals can be used to study the gravitational fields of the star and to test theories of gravity.

Neutron stars also have extremely strong magnetic fields, with strengths that can be as much as a billion times greater than the Earth's magnetic field. These magnetic fields can produce powerful bursts of radiation and can influence the behavior of matter in the surrounding environment.

The study of neutron stars is an important area of research in astrophysics, as it can provide insights into the behavior of matter under extreme conditions and can help us to better understand the nature of gravity and the structure of the universe.

## 2. CHARACTERISTICS OF NEUTRON STARS

Theoretical research on neutron stars has revealed several key characteristics of these extreme objects, including:

High Density: The matter in a neutron star is extremely dense, with a density that can be several times that of an atomic nucleus. The pressure inside the star is so high that it can sustain the gravitational force that would otherwise cause the star to collapse into a black hole (B.J. Schaefer & C. Fischer 2017).

Small Size: Despite their high mass, neutron stars have a very small size, with a diameter of only about 20 km. This extreme compactness is due to the strong gravitational forces that compress the matter in the star.

Strong Magnetic Fields: Neutron stars have incredibly strong magnetic fields, with strengths that can be as much as a billion times greater than the Earth's magnetic field. These magnetic fields can produce powerful bursts of radiation and can influence the behavior of matter in the surrounding environment (D. Vigano 2013).

Rapid Rotation: Neutron stars can rotate very rapidly, with some pulsars rotating hundreds of times per second. This rapid rotation is thought to be due to the conservation of angular momentum during the collapse of the star's core (A.C. Ordaz 2019).

Observational research on neutron stars has also provided important insights into their characteristics, including:

Pulsars: Pulsars are rapidly rotating neutron stars that emit beams of electromagnetic radiation. The precise timing of pulsar signals can be used to study the gravitational fields of the star and to test theories of gravity (P.T. Torres 2016).

X-Ray Emission: Neutron stars can emit X-rays due to the intense gravitational fields near their surface. Observations of X-ray emission can provide information about the structure and composition of the star's surface.

Gamma-Ray Bursts: Neutron stars can be involved in gamma-ray bursts, which are some of the most energetic events in the universe. Studying these bursts can provide information about the physics of the star's magnetic fields and the behavior of matter in extreme conditions (Z. Berezhiani & F. Frontera 2002).

Overall, theoretical and observational research on neutron stars has revealed a range of fascinating characteristics and properties, making these extreme objects a subject of ongoing interest and study in astrophysics.

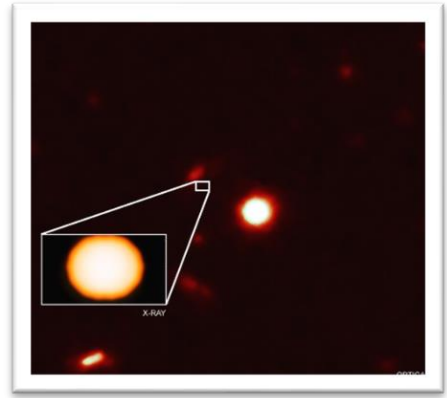

**Fig 1.** These images show the location of an event, discovered by NASA's Chandra X-ray Observatory, that likely signals the merger of two neutron stars. A bright burst of X-rays in this source, dubbed XT2, could give astronomers fresh insight into how neutron stars — dense stellar objects packed mainly with neutrons — are built.

Neutron stars are incredibly dense objects formed from the remnants of massive stars that have undergone supernova explosions. They are so dense that a single teaspoon of neutron star material would weigh about as much as a mountain on Earth.

The high density of neutron stars is due to the fact that their matter is compressed to extreme levels, with protons and electrons being squeezed together to form neutrons. This process is known as neutronization and results in a neutron-rich material with densities on the order of $10^{14}$ grams per cubic centimeter.

The exact density of a neutron star depends on several factors, including its mass and radius. However, it is generally believed that the average density of a neutron star is around $10^{17}$ kilograms per cubic meter, making them one of the densest objects in the known universe.

The extreme density of neutron stars is also responsible for their intense gravitational fields, which can warp space-time and cause gravitational lensing effects. These properties make neutron stars incredibly fascinating objects for astronomers and physicists to study, as they provide valuable insights into the nature of matter and the fundamental forces of the universe.

Neutron stars are incredibly small and compact, with typical radii on the order of only 10 kilometers (6.2 miles). This is due to their extreme density, which causes their mass to be packed into a very small volume.

To put the size of a neutron star into perspective, consider that the Sun, which is a relatively average-sized star, has a radius of approximately 700,000 kilometers (434,000 miles). This means that a neutron star is approximately 70,000 times smaller than the Sun, despite having a mass that is comparable to or greater than that of the Sun.

The small size of neutron stars also means that they have incredibly strong gravitational fields. In fact, their gravitational fields are so strong that they can cause time to slow down and space to warp in their vicinity, in a phenomenon known as gravitational time dilation.

Despite their small size, neutron stars are incredibly energetic objects and emit intense radiation across the electromagnetic spectrum, including X-rays and gamma rays. This radiation is thought to be generated by a variety of mechanisms, including the release of energy from the neutron star's magnetic field and the acceleration of particles in the neutron star's intense gravitational field.

The small size of neutron stars also makes them incredibly difficult to observe directly, but astronomers can detect their presence through a variety of indirect methods, such as observing the effects of their gravity on nearby objects or detecting the radiation they emit.

Neutron stars are known to possess incredibly strong magnetic fields, with strengths that can be trillions of times greater than the Earth's magnetic field. These magnetic fields are thought to be remnants of the magnetic fields that existed in the original star before it collapsed to form a neutron star.

The magnetic field of a neutron star is generated by the rotation of its charged particles, which causes electric currents to flow in the star's interior. These currents, in turn, generate a magnetic field that is aligned with the star's axis of rotation.

The magnetic fields of neutron stars are so strong that they can have a profound impact on the surrounding environment. For example, they can accelerate particles to extremely high energies, creating intense radiation that can be observed across the electromagnetic spectrum.

The strong magnetic fields of neutron stars can also cause them to emit beams of radiation from their magnetic poles, much like a lighthouse. These beams are typically observed as regular pulses of radiation, giving rise to the name "pulsars."

In addition to their impact on the surrounding environment, the magnetic fields of neutron stars are of great interest to physicists and astronomers because they provide valuable insights into the behavior of matter at extreme densities and temperatures. They also play an important role in shaping the structure and evolution of neutron stars over time.

Neutron stars are known for their incredibly rapid rotation, with some neutron stars spinning hundreds of times per second. This rapid rotation is a consequence of the conservation of angular momentum, which causes a collapsing star to spin faster and faster as it shrinks in size.

The fastest spinning neutron star currently known is PSR J1748-2446ad, which rotates at a rate of 716 times per second. This means that the neutron star completes one full rotation in just 1.4 milliseconds.

The rapid rotation of neutron stars has a number of important implications. For example, it causes the star's magnetic field to be compressed and amplified, leading to the production of intense radiation that can be observed across the electromagnetic spectrum. The rotation also causes the neutron star to flatten at the poles and bulge at the equator, creating an oblate shape that is sometimes likened to a spinning top.

The rapid rotation of neutron stars is also of great interest to physicists and astronomers because it provides insights into the nature of matter at extreme densities and temperatures. For example, the observation of pulsars, which are rapidly rotating neutron stars that emit beams of radiation, has provided important tests of Einstein's theory of general relativity and allowed astronomers to measure the properties of the interstellar medium.

Finally, the rapid rotation of neutron stars has important implications for the search for gravitational waves, which are ripples in the fabric of space-time predicted by Einstein's theory of general relativity. Rapidly rotating neutron stars are expected to emit gravitational waves, and their detection could provide important insights into the behavior of matter at extreme densities and temperatures.

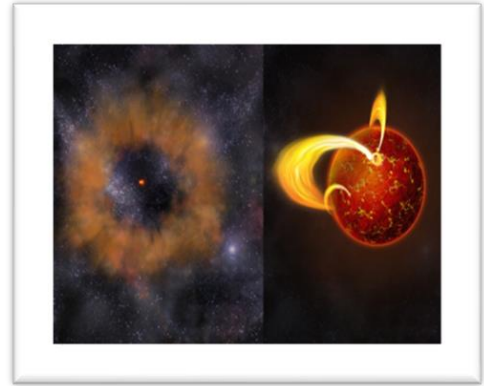

**Fig 2.** These illustrations show how an extremely rapidly rotating neutron star, which has formed from the collapse of a very massive star, can produce incredibly powerful magnetic fields. These objects are known as magnetars.

Neutron stars are extremely dense objects that are formed from the remnants of massive stars that have undergone a supernova explosion. They are composed almost entirely of neutrons and have a mass about 1.4 times that of the sun, but are only about 10 kilometers in diameter. Neutron stars are known for their strong magnetic fields and rapid rotation, which can lead to the emission of X-rays.

The intense magnetic fields of neutron stars can accelerate charged particles to very high energies, which in turn emit X-rays as they interact with the magnetic field. This process is known as synchrotron radiation. The X-rays emitted by neutron stars can be observed by telescopes in space, such as NASA's Chandra X-ray Observatory and the European Space Agency's XMM-Newton.

In addition to synchrotron radiation, neutron stars can also emit X-rays through a process known as thermal emission. As the neutron star cools down, it emits radiation at different wavelengths, including X-rays. The temperature of the neutron star's surface can be several million degrees, which is hot enough to emit X-rays.

X-ray emission from neutron stars can provide important information about their properties, such as their magnetic fields, temperatures, and compositions. X-ray observations can also help us to study the behavior of matter under extreme conditions, such as those found in neutron stars.

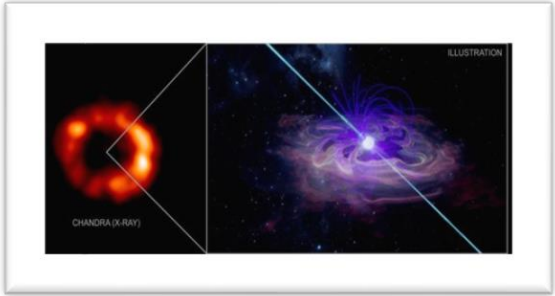

**Fig 3.** Supernova 1987A with a pulsar wind nebula at its center. Credit: Chandra (X-ray): NASA/CXC/Univ. di Palermo/E. Greco; Illustration: INAF-Observatory

Gamma-ray bursts (GRBs) are among the most energetic events known in the universe. They are short-lived bursts of gamma rays that can last anywhere from a fraction of a second to several minutes, and they release more energy in that brief period than the Sun will emit over its entire lifetime. Neutron stars are one of the possible sources of GRBs (A. Nathanael 2018).

Neutron stars are extremely dense objects that are formed when a massive star collapses under its own gravity. They are about the size of a city, but their mass is typically greater than that of the Sun. Neutron stars are also incredibly hot and can spin very rapidly, emitting intense radiation in the process.

GRBs are thought to be produced when two neutron stars collide or when a neutron star merges with a black hole. These collisions release an enormous amount of energy in the form of gamma rays, which are detected by telescopes in space. The gamma rays are produced as a result of the intense magnetic fields and shock waves that are generated during the collision.

Another possible source of GRBs from neutron stars is the so-called "magnetar" model. Magnetars are a type of neutron star that has an extremely strong magnetic field, billions of times stronger than that of ordinary neutron stars. These magnetic fields can store enormous amounts of energy, and when they are released, they can produce intense bursts of gamma rays.

In conclusion, neutron stars are one of the possible sources of gamma-ray bursts, and their collisions or magnetic field eruptions can release enormous amounts of energy in the form of gamma rays. Studying these events can help us better understand the behavior of matter under extreme conditions and the physics of the early universe.

## 3. PROPERTIES OF NEUTRON STARS: OBSERVATIONAL RESEARCH

Observational research has revealed many fascinating properties of neutron stars. Here are a few key findings:

Mass and size: Neutron stars are extremely dense, with masses typically ranging from 1.4 to 2.1 times that of the Sun, but their radii are only around 10-15 kilometers. This means that their densities can be as high as several times that of atomic nuclei.

Rotation: Many neutron stars are rapidly rotating, with some spinning hundreds of times per second. This rapid rotation is thought to be due to the conservation of angular momentum during their formation. The fastest known neutron star is PSR J1748-2446ad, which rotates at a rate of 716 times per second.

Pulsars: Neutron stars can emit beams of radiation that sweep across the sky like a lighthouse beam. These objects are known as pulsars, and they were first discovered in the 1960s. Pulsars are thought to be neutron stars that are oriented in such a way that their radiation beams can be observed on Earth.

Magnetic fields: Neutron stars can have extremely strong magnetic fields, with strengths that can be trillions of times that of the Earth's magnetic field. These magnetic fields can create powerful electromagnetic radiation and can affect the behavior of matter in the star's vicinity.

Neutron star mergers: Observations of gravitational waves have revealed that neutron star mergers can occur, and these events can produce a variety of interesting phenomena, including gamma-ray bursts, kilonovae, and r-process nucleosynthesis.

Neutron star crusts: Neutron stars are thought to have a crust made up of solid nuclear matter. Observations of x-ray bursts and oscillations in the x-ray flux from neutron stars have provided important insights into the properties of these crusts, including their composition and thickness.

Observational research has been essential in advancing our understanding of neutron stars and their properties. Further observations and analyses of these fascinating objects will undoubtedly continue to shed light on some of the most extreme and enigmatic phenomena in the universe.

Neutron stars are fascinating astronomical objects that are incredibly dense and have very strong gravitational fields. Here are some observations that have been made of neutron stars:

Pulsars: Neutron stars that emit beams of electromagnetic radiation from their poles and appear to pulse as they rotate. These pulsating signals have been observed in the radio, X-ray, and gamma-ray bands of the electromagnetic spectrum.

X-ray Bursts: Neutron stars in binary systems can accumulate matter from their companion star, which can lead to X-ray bursts when the accumulated matter ignites on the surface of the neutron star.

Gravitational Waves: The merger of two neutron stars was detected in 2017 through the detection of gravitational waves. This event, called GW170817, was also observed through electromagnetic radiation in the form of gamma-rays, providing a multi-messenger view of the event.

Magnetars: Neutron stars with extremely strong magnetic fields, which can produce intense bursts of X-rays and gamma-rays.

Accretion Disks: Some neutron stars in binary systems can have accretion disks, which are formed by the accumulation of matter from their companion star. These disks can emit X-rays and other forms of electromagnetic radiation.

Neutron Star Crust: Observations of the surface of neutron stars can provide insight into the composition and structure of the neutron star crust, which is thought to be made up of a lattice of atomic nuclei immersed in a sea of electrons.

Neutron Star Cooling: The temperature of neutron stars can be measured by observing their thermal radiation. By studying the cooling of neutron stars over time, astronomers can learn about their internal structure and the properties of ultra-dense matter.

## 4. PULSARS

Pulsars are a type of neutron star that emits beams of electromagnetic radiation from their magnetic poles. These beams are often observed as regular pulses of radiation, hence the name pulsars. The pulsations are due to the rotation of the neutron star, which causes the beams of radiation to sweep across our line of sight like a lighthouse beam.

Pulsars were first discovered in 1967 by Jocelyn Bell Burnell and Anthony Hewish. Since then, thousands of pulsars have been discovered, mainly through radio astronomy observations. Pulsars are also observed in other parts of the electromagnetic spectrum, including X-rays and gamma-rays (R. Grootjans 2016).

Pulsars are extremely precise timekeepers, with some pulsars having rotation periods as short as a few milliseconds. They also have incredibly strong magnetic fields, which can be up to a billion times stronger than the magnetic field of the Earth. These magnetic fields can produce a variety of interesting phenomena, such as accelerating particles to very high energies and producing intense X-ray and gamma-ray emission.

Pulsars are important objects for studying a variety of astrophysical phenomena, including gravitational waves, general relativity, and the properties of ultra-dense matter. They also provide a unique laboratory for studying extreme physics and can help us understand the evolution and structure of neutron stars.

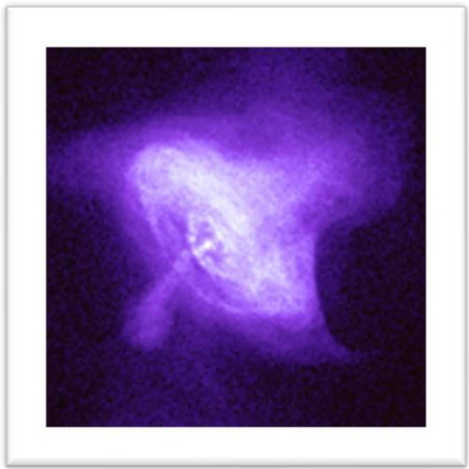

**Fig 4.** The Crab Nebula is the remnant of a supernova explosion that was seen on Earth in 1054 AD. It is 6000 light years from Earth. At the center of the bright nebula is a rapidly spinning neutron star, or pulsar that emits pulses of radiation 30 times a second.

## 5. GRAVITATIONAL WAVES IN NEUTRON STARS

Gravitational waves are ripples in the fabric of spacetime that are generated by accelerating masses. Neutron stars, being incredibly dense objects with strong gravitational fields, are prime sources of gravitational waves. Here are some ways in which gravitational waves can be generated by neutron stars:

Binary Neutron Star Mergers: When two neutron stars orbit each other in a binary system, they emit gravitational waves as they spiral towards each other due to the emission of energy in the form of gravitational waves. The waves become stronger and more frequent as the stars get closer together, leading to a rapid increase in the strength of the signal shortly before the merger. The detection of gravitational waves from the merger of two neutron stars, called GW170817, was a landmark event in gravitational wave astronomy.

Asymmetric Rotations: Neutron stars that have an irregular or asymmetric shape, or that experience rapid changes in their rotation rate, can also emit gravitational waves. These gravitational waves are typically in the millisecond range and can be observed by ground-based detectors such as LIGO and Virgo (F. Gittins 2021).

Non-Radial Oscillations: Neutron stars can also undergo non-radial oscillations due to disturbances such as starquakes or accretion of material. These oscillations can also produce gravitational waves, which can be observed through their effects on the pulsation frequencies of the star.

The detection of gravitational waves from neutron stars has opened up a new window on the universe, allowing us to study the properties of ultra-dense matter, test general relativity in extreme environments, and explore the formation and evolution of binary star systems.

## 6. MAGNETARS

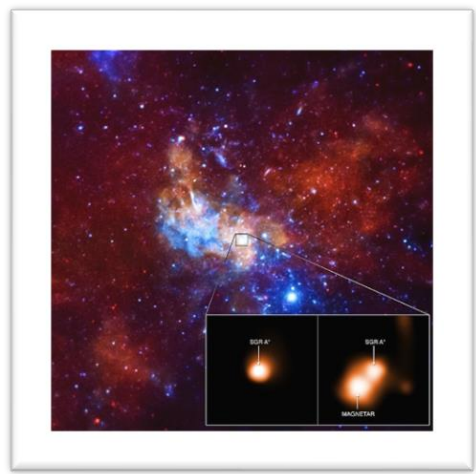

**Fig 5.** SGR 1745-2900: Magnetar Near Supermassive Black Hole Delivers Surprises.

Magnetars are a type of neutron star that have extremely strong magnetic fields, with strengths up to a billion times stronger than those of typical neutron stars. These magnetic fields are so strong that they can affect the behavior of matter on the surface of the star and in its immediate surroundings (M. B. Hoven 2012).

Magnetars were first proposed as a distinct class of objects in 1992, and the first confirmed magnetar was discovered in 1998. They are thought to be the result of the collapse of a massive star, which leaves behind a highly magnetized neutron star. Magnetars emit high-energy electromagnetic radiation, including X-rays and gamma rays, and are often observed to undergo periodic outbursts of intense activity.

The intense magnetic fields of magnetars are believed to give rise to a variety of exotic phenomena, including the emission of high-energy photons, the production of ultrastrong electric fields, and the acceleration of particles to nearly the speed of light. These properties make magnetars important objects of study for astrophysicists, as they provide unique opportunities to probe the behavior of matter and radiation under extreme conditions.

## 7. ACCRETION DISKS

An accretion disk is a structure that forms around a compact object, such as a black hole or a neutron star, when it is pulling in matter from a companion star or from the surrounding interstellar medium. As the matter falls towards the central object, it forms a disk-like structure, which can become very hot and emit significant amounts of radiation.

Accretion disks are an important phenomenon in astrophysics, as they play a key role in a number of astrophysical processes. For example, they can generate intense X-ray and gamma-ray radiation, and they are thought to be responsible for the jets of high-energy particles that are observed emanating from many active galaxies and quasars.

The structure and behavior of an accretion disk depend on a number of factors, including the mass and spin of the central object, the properties of the matter being accreted, and the rate at which matter is flowing into the disk. Studying the properties of accretion disks can therefore provide important insights into the underlying physics of compact objects and the processes that occur in the most extreme environments in the universe (K.M. Kratter 2010).

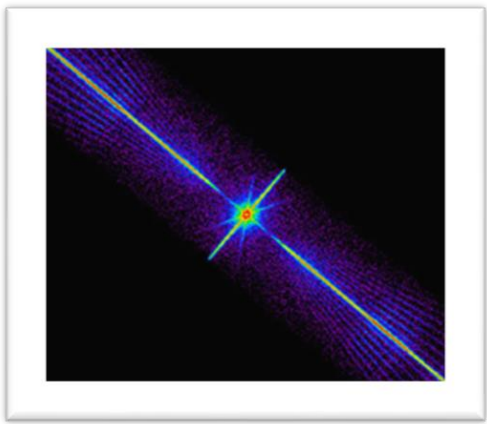

**Fig 6.** Chandra Pinpoints Edge of Accretion Disk Around Black Hole

This Chandra X-ray Observatory image is a spectrum of a black hole, which is similar to the colorful spectrum of sunlight produced by a prism. The X-rays of interest are shown here recorded in the bright stripe that runs rightward and leftward from the center of the image. These X-rays are sorted precisely according to their energy with the highest-energy X-rays near the center of the image and the lower-energy X-rays farther out. The spectrum was obtained by using the Low Energy Transmission Grating (LETG), which intercepts X-rays and changes their direction by amounts that depend sensitively on the X-ray energy. The LETG is activated by swinging an assembly into position behind the mirrors and in front of the instrument that detects the X-rays. The assembly holds 540 gold transmission gratings; when in place behind the mirrors, the gratings intercept the X-rays reflected from the telescope. The bright spot at the center is due to a fraction of the X-ray radiation that is not deflected by the LETG. The spokes that intersect the central spot and the faint diagonal rays that flank the spectrum itself are artifacts due to the structure that supports the LETG grating elements

A team of scientists led by Jeffrey McClintock (Harvard-Smithsonian Center for Astrophysics) used the LETG in conjunction with the Advanced CCD Imaging Spectrometer (ACIS) detector to observe the black hole binary system known as XTE J1118+480 for 27,000 seconds on April 18, 2000. This "X-ray nova," so-called because it undergoes occasional eruptions followed by long periods of dormancy, contains a Sun-like star orbiting a black hole.

## 8. NEUTRON STAR COOLING

Neutron stars are incredibly dense and compact objects that are created when a massive star explodes in a supernova. They are composed almost entirely of neutrons and have a mass of about 1.4 times that of the sun, but a radius of only about 10 km (A.Y. Potekhin 2015).

Neutron stars are extremely hot when they are first formed, with temperatures reaching up to billions of degrees. However, over time they cool down through various processes, including thermal radiation, neutrino emission, and the formation of a solid crust.

One of the primary cooling mechanisms for a neutron star is thermal radiation. As the star cools, it emits electromagnetic radiation in the form of X-rays, gamma rays, and visible light. This radiation carries away energy from the star, causing it to cool further.

Another important cooling mechanism is neutrino emission. Neutrinos are subatomic particles that are produced in the core of the neutron star through various nuclear reactions. They escape the star easily and carry away energy, causing the star to cool.

The formation of a solid crust also plays a role in cooling. As the neutron star cools, the outer layers of the star solidify, forming a crust. This process releases energy and further cools the star.

The rate of cooling of a neutron star depends on a number of factors, including its mass, composition, and magnetic field. Observations of cooling neutron stars can provide valuable insights into the physics of extremely dense matter and the behavior of fundamental particles such as neutrinos.

## RESULTS AND DISCUSSION

Neutron stars are fascinating objects that are the result of the collapse of massive stars. They are incredibly dense, with a mass comparable to that of the sun but a radius of only about 10 kilometers. In this discussion, we will explore the theoretical and observational research on the structural characteristics and physical properties of neutron stars.

Structural Characteristics:

The structure of a neutron star is determined by the properties of its constituents, which include neutrons, protons, electrons, and possibly other exotic particles. The central region of a neutron star is believed to be composed primarily of neutrons, which are held together by the strong nuclear force. The outer regions of the neutron star may contain a mixture of protons, electrons, and other particles.

The equation of state (EOS) of matter inside neutron stars is a crucial parameter that determines the structure and properties of these objects. The EOS describes the relationship between the pressure, density, and temperature of the matter inside the neutron star. The theoretical EOSs of neutron star matter have been constructed using a variety of models, including nuclear physics calculations, lattice QCD simulations, and effective field theories.

Observations of neutron star masses and radii provide important constraints on the EOS of neutron star matter. The masses of neutron stars can be measured using pulsar timing observations, while the radii can be inferred from the analysis of X-ray spectra emitted from their surfaces. Recent observations have provided strong evidence for relatively large neutron star radii, which suggest that the EOS of neutron star matter may be relatively soft.

Physical Properties:

Neutron stars have a number of physical properties that are of great interest to astrophysicists, including their surface temperature, magnetic field strength, and rotational properties.

The surface temperature of a neutron star is determined by the balance between the energy emitted by the star and the energy produced by the nuclear reactions in its interior. Neutron stars are observed to have surface temperatures ranging from a few hundred thousand to a few million Kelvin, depending on their age and other factors.

The magnetic field strength of neutron stars is also of great interest. Neutron stars are believed to have extremely strong magnetic fields, which can reach values of up to $10^{15}$ Gauss. These fields can produce a variety of interesting phenomena, including pulsar emission, magnetar flares, and particle acceleration.

The rotational properties of neutron stars are also of great interest. Neutron stars can rotate at incredibly high speeds, with some pulsars rotating hundreds of times per second. The rotational energy of a neutron star can be tapped into by accreting matter from a companion star, producing X-ray emission and other observable phenomena.

In conclusion, the study of neutron stars is a rapidly evolving field of research that combines theoretical modeling with observations across a range of wavelengths. Advances in our understanding of the structural characteristics and physical properties of neutron stars will continue to be driven by a combination of theoretical and observational research.

## CONCLUSION

In conclusion, neutron stars are fascinating objects with unique structural characteristics and physical properties. Theoretical models and observational studies have provided significant insights into the nature of these objects, including their internal composition, equation of state, surface temperature, magnetic field strength, and rotational properties.

The study of neutron stars continues to be a vibrant area of research, with ongoing efforts to refine theoretical models and develop new observational techniques. Recent advances in gravitational wave astronomy, for example, have opened up new avenues for studying neutron star mergers and their aftermaths.

Ultimately, a more complete understanding of the structural characteristics and physical properties of neutron stars will not only deepen our knowledge of these objects but also provide insights into fundamental physics and the evolution of the universe as a whole.